# Controlled growth of $CH_3NH_3PbI_3$ nanowires in arrays of open nanofluidic channels

Massimo Spina[1], Eric Bonvin[1], Andrzej Sienkiewicz[1,2], Bálint Náfrádi[1], László Forró[1] & Endre Horváth[1]

Spatial positioning of nanocrystal building blocks on a solid surface is a prerequisite for assembling individual nanoparticles into functional devices. Here, we report on the graphoepitaxial liquid-solid growth of nanowires of the photovoltaic compound $CH_3NH_3PbI_3$ in open nanofluidic channels. The guided growth, visualized in real-time with a simple optical microscope, undergoes through a metastable solvatomorph formation in polar aprotic solvents. The presently discovered crystallization leads to the fabrication of $mm^2$-sized surfaces composed of perovskite nanowires having controlled sizes, cross-sectional shapes, aspect ratios and orientation which have not been achieved thus far by other deposition methods. The automation of this general strategy paves the way towards fabrication of wafer-scale perovskite nanowire thin films well-suited for various optoelectronic devices, e.g. solar cells, lasers, light-emitting diodes and photodetectors.

One-dimensional nanostructures, such as nanowires and nanotubes, represent the smallest dimension for efficient transport of electrons and excitons[1]. In particular, organic or solid-state nanowires, due to their unique properties as compared to their bulk counterparts, are attractive building blocks in a wide range of applications, including sensors[2,3], nanoelectronics[4,5], photonics[6,7], and renewable energy[8]. Recently, a solvatomorph-mediated synthesis of hybrid halide perovskite in nanowire form, specifically methylammonium lead iodide, $CH_3NH_3PbI_3$ (hereafter $MAPbI_3$), a material that has attracted overwhelming attention, was discovered[9–12]. Even if very little is known concerning their liquid phase growth mechanism or their structural and photo-physical properties, elongated lead halide perovskite particles[9,13] have already been successfully integrated into perovskite-based solar cells[14] and ultrasensitive, micro-fabricated photodetectors[15]. Recently, exceptional lasing properties of solution-grown perovskite thin films[16,17]; and nanowires[18] have been reported.

Solution-based low-temperature processes are generally recognized to be cost-effective and easy to scale-up fabrication methods. Therefore, the forecasted low fabrication cost is one of the main arguments of researchers and investors for considering organometallic halide perovskites as a promising material for the development of next-generation photovoltaics[19]. Despite numerous successful reports on record-breaking efficiencies it remains a major challenge to attain device-to-device reproducibility of the performance parameters, even in case of lab-scale (~20 $mm^2$ surface area) prototypes[20]. Currently, the major hurdle to overcome is related to the lack of precise control of the crystal parameters, such as the crystal habit, crystallite size, crystallinity and the quality of grain boundaries on small and large scale surfaces. For instance, in polycrystalline nanoparticle-based perovskite films, the most important parameters to achieve outstanding optoelectronic performances are strongly linked to both the size and the orientation of the crystalline domains, likewise affecting the film homogeneity, area coverage, pinholes and roughness[21]. Similarly, the slip-coating method, *i.e.* our previously reported single-step approach, produces perovskite nanowires in dandelion-like organization on a flat surface.

The liquid phase growth pattern is determined by the formation of nucleation centers randomly distributed over the substrate, therefore there is little control on the area coverage, pinholes, aspect ratio and orientation of the nanowires[9,15]. Recently, however, we found a fairly simple approach to overcome the spatially-random surface nucleation. Therefore, here, we report on the guided growth of extremely high aspect-ratio perovskite crystalline nanowires in the arrays of open nanofluidic channels. An important feature of the observed "solvatomorph-graphoepitaxy" is that the crystallization of $MAPbI_3$ solvatomorphs proceeds exclusively in polar aprotic solvents such as dimethylformamide (DMF), dimethylacetamide (DMAc) and dimethyl sulfoxide (DMSO), at near room temperature. Moreover, for the first time, the kinetics of the nanowire growth was visualized with a simple optical microscope in real-time. We envision that standardization of this general strategy will allow the reproducible fabrication of wafer-scale perovskite nanowire thin films with highly controlled crystallite dimensions and crystallinity. Their integration into various optoelectronic devices could help in further boosting their performances.

[1]Laboratory of Physics of Complex Matter (LPMC), Ecole Polytechnique Fédérale de Lausanne, 1015 Lausanne, Switzerland. [2]ADSresonances, CH-1028, Préverenges, Switzerland. Correspondence and requests for materials should be addressed to E.H. (email: endre.horvath@epfl.ch)

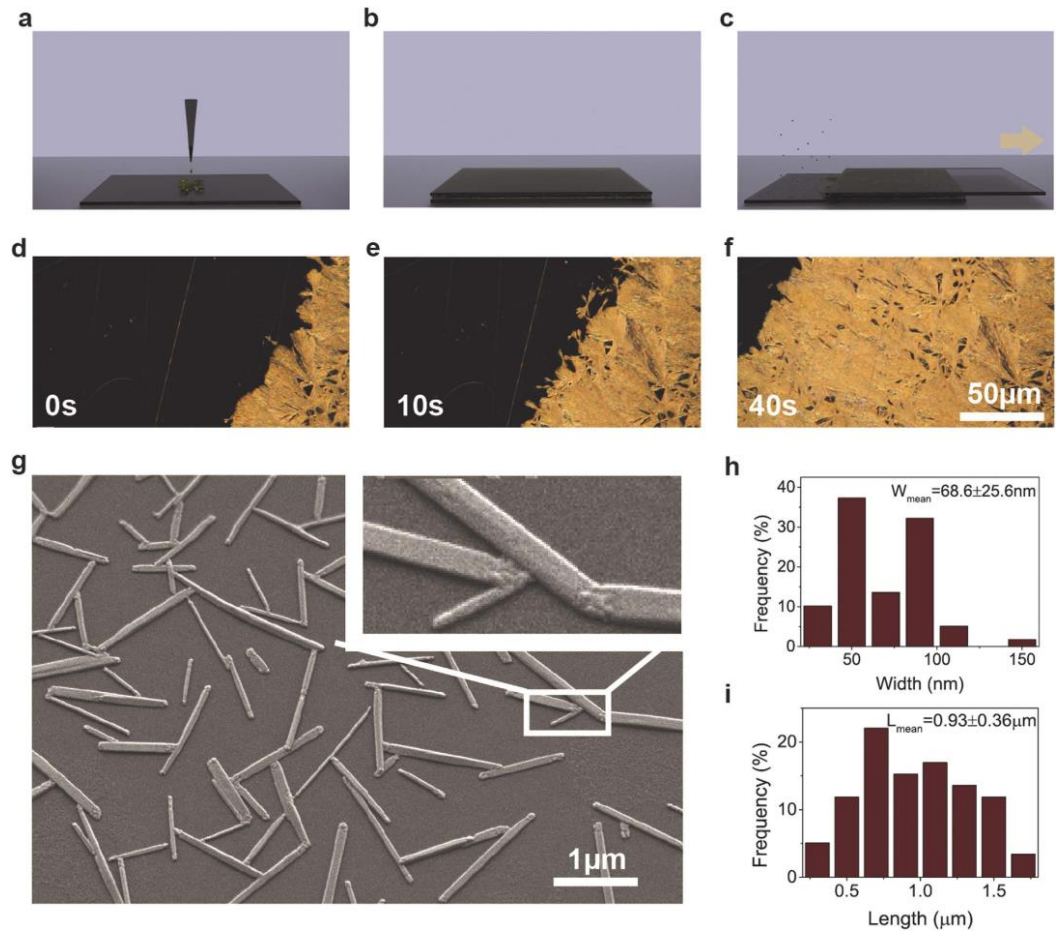

**Figure 1. Schematic illustration of the slip-coating process** (**a–c**) and optical images extracted from a video of the growth of a set of slip-coated MAPbI$_3$ nanowires (**d–f**). SEM micrographs of a representative set of nanowires (**g**). AFM distribution of length and width of the set of nanowires shown in g. The liquid phase growth pattern is inherently determined by the formation of nucleation centers randomly distributed on the surface. The nanowires grow in dandelion-like pattern. The nanowires grow in radial direction out from the nucleation centers and form an assembly of crystallites that resembles a "paper fan". We do see wire-wire intergrowth but we do not observe the formation of secondary bunches. The slip-coating process has no precise control over the dimensions of the as-synthesized nanowires.

We show that slip-coating of saturated solution of MAPbI$_3$ in polar aprotic solvents (i.e. DMF, DMSO and DMAc) results in the formation of perovskite nanowires organized in dandelion-like arrangement on a silicon surface. As can be seen from the optical microscopy and scanning electron microscopy (SEM) images in Fig. 1, the liquid phase growth pattern is inherently determined by the formation of nucleation centers, which are randomly distributed on the surface. Therefore, there is little control over the aspect ratio and the orientation of the nanowires. Only a very slight, shear-force-induced guidance was observed with respect to the sliding direction of the top glass plate.

MAPbI$_3$ has been reported to be unstable in the majority of ordinary solvents, including water, and to decompose at moderate temperatures (~400 K) in addition[22,23]. Therefore, processing and precise patterning by post-growth assembly techniques such as dielectrophoresis[24], Langmuir-Blodgett self-assembly[25] and mechanical shear[1] seems to be a challenging task. Unlike the above mentioned multi-step post-growth assembly techniques, here, we combine a bottom-up nanofluidic alignment with a top-down surface patterning technique to achieve both the synthesis and oriented assembly of elongated MAPbI$_3$ crystallites in a single step. The nanowires were crystallized in the arrays of nanofabricated channels owing to the strong guiding effect of the nano-grooves[26]. Similar guiding effects have already been observed for chemical vapor deposited (vapor-liquid-solid grown, VLS) grown gallium nitride (GaN) nanowires[27]. The phenomenon of the alignment induced by the intimate epitaxial relationship has been discovered four decades ago and it has been named "graphoepitaxy"[28–30]. In a typical experiment, we dropped a supersaturated solution of $MAPbI_3$ dissolved in DMF onto the arrays of nanostructured trenches etched in a $SiO_2$ substrate (Fig. 2a). Next, capillary forces drove the liquid inside the channels with a speed proportional to the channel width[31] (Fig. 2b). At this point, the first defects and etching-induced crystallographic imperfections of the channel wall triggered the heterogeneous nucleation of a yellow, so far barely studied clathrate phase of $MAPbI_3$-DMF. This crystalline, translucent-yellow precipitate contained the mother liquor, that is a polar aprotic solvent (*e.g.* DMF, DMSO and DMAc). Therefore, it can be regarded as a solvatomorph phase (*i.e.* metastable precursor phase) of the $MAPbI_3$ perovskite. Unlike $MAPbI_3$ itself, this solvatomorph phase does not show the characteristic photo-luminescence under excitation with the green monochromatic incoherent light ($\lambda_{ex}$ = 546 nm) (Fig. S4). Interestingly, neither the solvatomorph formation nor the guided growth was observed with gamma-Butyrolactone (GBL), another commonly used solvent in solution based perovskite thin films preparation. This partially explains the enigma of the highly-anisotropic crystallization of cubic or tetragonal phases of the perovskite. Actually, the anisotropic crystal growth is specific to the metastable translucent-yellow colored solvatomorph phase, formed by the host-guest interaction (solvatomorph formation) of $MAPbI_3$ with polar aprotic solvents ($MAPbI_3$-DMF, $MAPbI_3$-DMSO and $MAPbI_3$-DMAc). It is important to realize, that these are metastable precursor compounds, hence the final $MAPbI_3$ perovskite phase is formed by a subsequent solvent evaporation-induced recrystallization of the solvatomorph phase, which is equivalent to the dehydration processes observed for instance in many oxo-hydroxo compounds (Fig. 2g,h). The analysis of the time-lapse videos recorded in an optical microscope provided insight into the nanowire formation mechanism and also added another simple tool to study the kinetics of growth and dissolution (Fig. 2, Fig. S6, video S1). We have found, that the crystallization follows the classical solvent evaporation-induced supersaturation-driven crystallization. The most important parameters controlling the growth rate are the surface-normalized concentration of the $MAPbI_3$ solution, as well as the temperature and the surface tension of the solvent. In the first approximation, stable clusters and nuclei of MAPbI$_3$-DMF solvatomorph form in the channel entrances, thus acting as foreign particles (Fig. 2e). The subsequent nanowire growth, dilutes the liquid, hence shifts the solution to an undersaturated condition (Fig. 2f). This is expected to slow down the kinetics and ultimately stop the precipitation of the solute. However, the solvent evaporation from the open nanofluidic channel acts on the opposite way, it concentrates the solution, maintaining the supersaturation condition in equilibrium (Fig. 2g). Furthermore, the capillary forces as well as the concentration difference-induced diffusion of the solute toward the growing crystal plane, *i.e.* the growing end of the wire, play a fundamental role in crystal growth. When the continuous supply of the solute and/or the mother liquor is blocked the supersaturation is exhausted, the nanowire growth stops (Fig. 2c). The synthesis process ends when all the mother liquor escapes from the metastable clathrate phase (MAPbI$_3$-DMF) formed by the polar aprotic solvent. Nanoscale crystallite dimensions allow the solvent to escape without inducing significant shrinkage-induced cracks or damage in the crystal habit. Ultimately, the precursor phase transforms to the final, grey-silver MAPbI$_3$ phase and maintaining the elongated crystal shape (Fig. 2d).

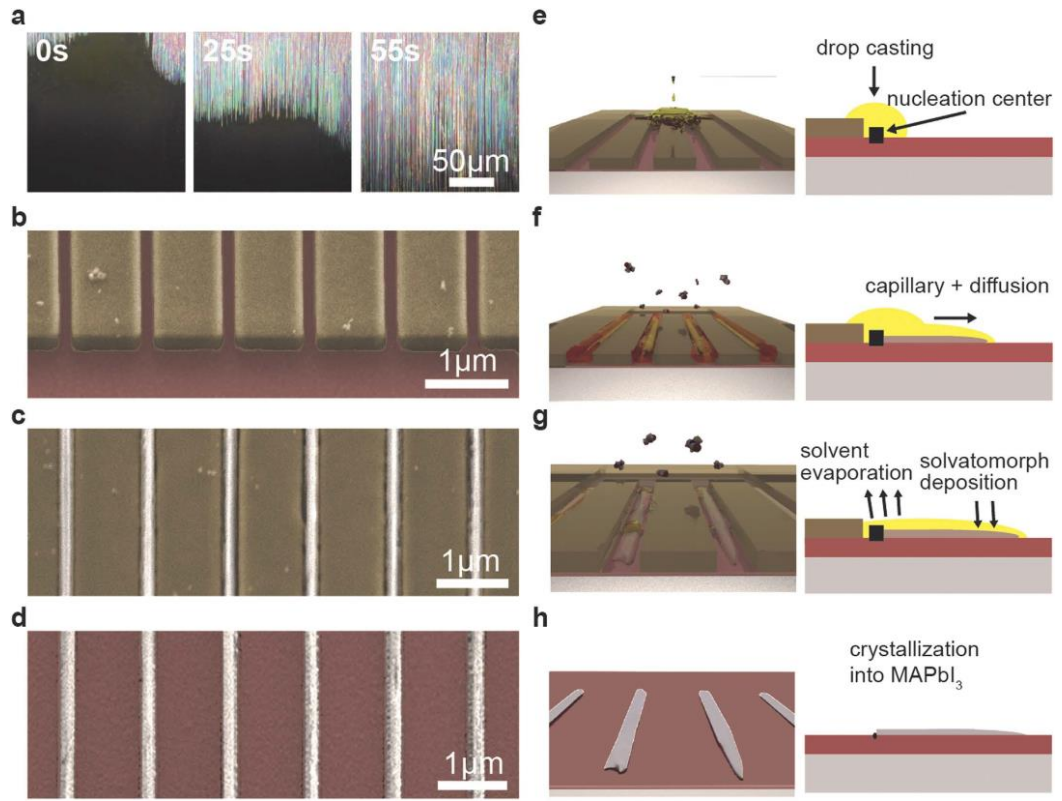

**Figure 2. Snapshots of a video showing the graphoepitaxial growth process in a dense array of 500 nm wide channels etched in Si (a)**. False-color SEM micrographs of the growth process implemented in nanofluidic channels realized with a high resolution positive e-beam resist, ZEP520A (**b–d**). 200 nm-wide array of ZEP520A nanochannels on $SiO_2$ substrate (**b**) $MAPbI_3$ nanowires synthesized in the nanofluidic channels (**c**) $MAPbI_3$ nanowires after the removal of ZEP520A resist by chloroform (**d**). Conceptual illustration of the aligned growth of $MAPbI_3$ nanowires in nanofluidic channels (**e–h**) The saturated $MAPbI_3$ solution is drop-casted on a series of open nanofluidic channels realized by ZEP520A (**e**). The solution is driven by capillary forces inside the channels. The nucleation takes place at the first defect present in the channel and the growth process starts (**f**). When the solution supply is stopped, the system does not fulfil the growth conditions anymore and synthesis stops (**g**). After all the solvent has evaporated, the nanowires transform from their metastable phase to $MAPbI_3$ (**h**). Detailed structural and photophysical characterization of $MAPbI_3$ nanowires have been reported in our previous work.[9,15]

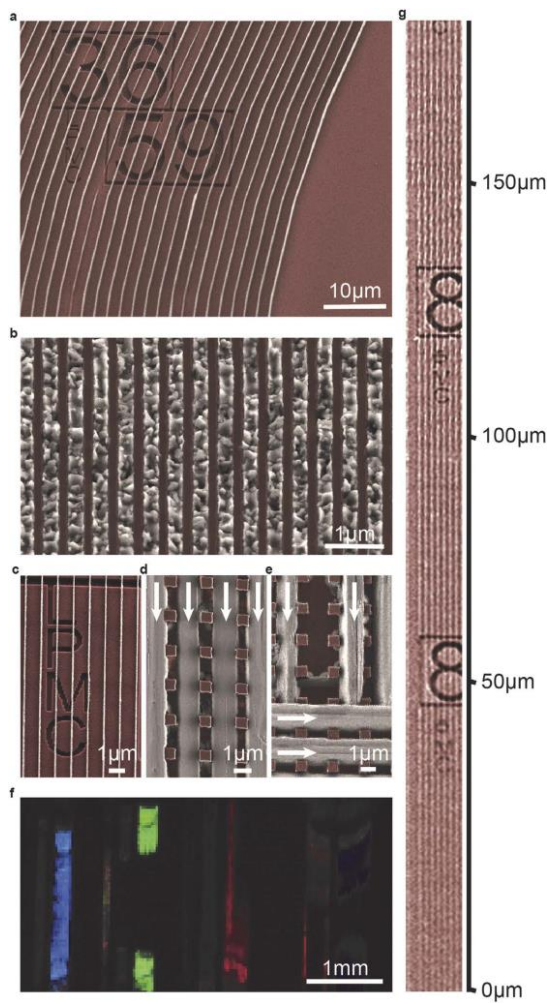

**Figure 3. Colored SEM micrographs of different set of MAPbI$_3$ nanowires.** Curved nanowires after ZEP520A removal (**a**) dense array of aligned MAPbI$_3$ nanoparticles (**b**) nanowires over etched patterns (**c**) perovskite nanowires grown with the guidance of Si pillars (**d**) that can be used to obtain cross-bar architectures; the growth directions are marked with arrows (**e**) Optical image of mm$^2$-sized surfaces composed of MAPbI$_3$ nanowires with different widths and spacings illuminated with white light. The blue, green, orange and red colors are due to the interference pattern of millimeter long periodic nanowire arrays (f and Video S2); extremely long, 200 nm-wide array of MAPbI$_3$ nanowires on SiO$_2$ after ZEP520A removal (**g**).

The process yields nanowires with well-defined sizes, cross-sections and spatial distributions. With this fairly simple technique, extremely long (up to few mm) and narrow (down to 10 nm) V-shaped (Fig. S2a) and rectangular-shaped cross section (Fig. S2b) nanowires have been realized along predefined nanofluidic channels. Examination of the growth process on larger length scales shows that the growth extends over millimeter length scales and seems to be limited by the length of the fluidic channel. We have carried out several experiments in order to understand the parameters controlling the crystal habit.

We have found that whereas wider, micrometer-sized channels were mostly filled with bundles of nanowires (Fig. S2d), individual nanowires tend to grow in narrower channels. This can be explained by the stochastic nature of the heterogeneous nucleation process. Narrow channels are comparable in size to the first nucleus formed, hence there is a higher probability that they contain less nucleation centers, resulting in individual elongated single crystallites (Fig. S1). In contrast, wider channels (>1 μm) contain more imperfections, where multiple nuclei can be formed, resulting in the formation of crystallographically-fused parallel aggregates of perovskite nanowires (Fig. S2d, Video S4). The crystal growth of the elongated solvatomorphs is likewise feasible in curved or zig-zag shaped nanofluidic channels (Fig. 3a). The graphoepitaxial growth is not limited to a particular inorganic substrate, hence the guided growth was similarly observed in metallic (e.g Au) as well as in organic resin (e.g. ZEP520A) based nanofluidic channels. This compositional versatility enables the fabrication of numerous patterned substrate/nanowire material combinations i.e. p-i-n lateral interdigital configurations (Fig. 3). The possible advantage of these non-conventional p-i-n architectures over the *state-of-the-art* planar or mesoscopic configurations is yet to be determined (Fig. S7). Interestingly, we observed guiding effects even in between non- continuous -walled channels, e.g. arrays of rectangular pillars of SiO$_2$ or organic resin of ZEP520A (Fig. 3d,e). This implicates that the most important parameter determining the nanowire diameter might be the size of the nucleation center. Once the nanowire growth starts, the cross-section remains constant over ~cm length scales (Fig. 3g, Video S3, Video S4). This allowed us to create controlled 'wire-to-wire' connections by launching the wire growth on the pillar-patterned silicon surface from two directions (90° degree angle) simultaneously (Fig. 3e). In addition, the surface coverage and the film thickness can be similarly controlled by the nanofluidic channel dimensions. Importantly, all the wires were spatially confined and the minimum separation of the aligned nanowires depends solely on the resolution and precision of the applied lithographical process. By using recent electron-beam tools, this can easily be reduced below 50 nm. Accordingly, the channel width and its periodicity will control the area coverage of a given film, with the film thickness being linearly proportional to the channel height.

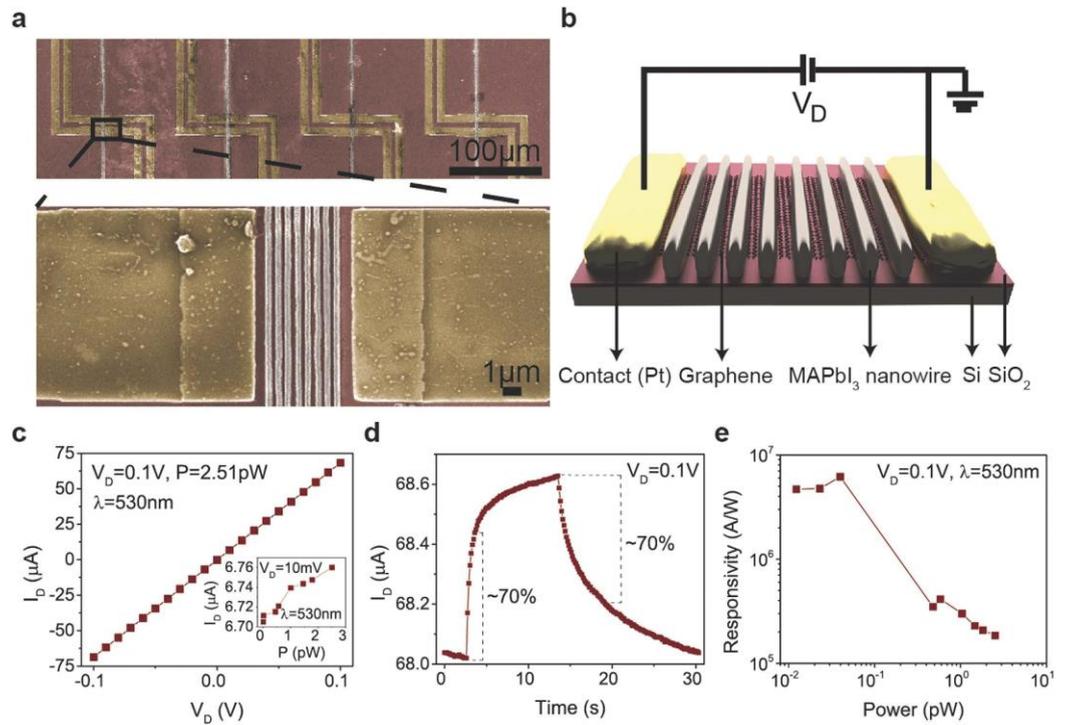

**Figure 4. Colored SEM micrographs of four parallel graphene/MAPbI$_3$ nanowires photodetector devices (upper panel) and a typical device composed of eight MAPbI$_3$ nanowires (grey color) grown on graphene contacted with platinum electrodes.** The nanowires were grown in 250 nm wide ZEP520A channels. The ZEP520A resist was dissolved prior the SEM imaging (**a**). Diagram of the device architecture (**b**) I-V curve of the best performing MAPbI$_3$ nanowire sensitized graphene photodetector illuminated with pW light intensities (**c**) Time response of a representative device (**d**). Two regimes can be identified: a fast one corresponding to ~70% decay (~2–5 s) and a slow one of ~10–20 s associated with the charge traps in nanowire film. Responsivity of the device for different incident light intensities (**e**).

As an example, we fabricated series of perovskite nanowire based photodetector devices using e-beam lithography (Fig. 4e). The longitudinal composite devices were composed of lithographically patterned graphene sensitized by a perovskite nanowire active layer (Fig. 4a,b). Recently we have reported on the fabrication of similar devices attaining remarkable responsivities ($2.6 \times 10^6$ AW$^{-1}$) under pW light intensities. Here, the hybrid devices prepared by the graphoepitaxial liquid-solid growth of aligned, millimeter-long nanowires of MAPbI$_3$ in open nanofluidic channels showed similar linear *I-V* characteristics, with characteristic response times of 2–5 seconds (Fig. 4d). The best devices reached responsivities as high as $6 \times 10^6$ AW$^{-1}$ (Fig. 4e). The drastic enhancement of the responsivity at very low light intensities (pW) could enable MAPbI$_3$ nanowire/graphene devices for use as low-light imaging sensors and single photon detectors. We attribute these very high device performances mainly to the controlled growth of perovskite nanowires in predefined positions. Our results demonstrate an important step in the integration of perovskite nanowires in arrays of microfabricated optoelectronic devices with high reproducibility. The process allows the synthesis of extremely long (~cm) and thin (~few nm) nanowires with a morphology defined by the shape of nanostructured open fluidic channels. Moreover, optimized spacing, shape and size in the periodic pattern may result in large-area, wafer-scale superstructures with advantageous optical properties for photonic, optoelectronic, radio-frequency and non-linear optic applications. Ultimately, the graphoepitaxial nanowire growth enables well-engineered angular restriction designs, which can improve the photon reabsorption process and thus maximize the performance of perovskite nanowire based optoelectronic device[32]. This low temperature solution growth method, unique in its gender opens an entirely new spectrum of architectural design of organometal halide perovskite based optoelectronic devices.

## Acknowledgements

This work was partially supported by the Swiss National Science Foundation and ERC ADVANCED GRANT (PICOPROP#670918). Device fabrication was carried out in part in the EPFL Center for Micro/Nanotechnology (CMI).

## Author Contributions

L.F. initiated the research. E.H. synthesized the perovskite solutions and discovered the graphoepitaxial nanowire growth. M.S. and E.B. prepared the microfabricated devices. M.S., E.B. and B.N. performed the photocurrent measurements and analyzed the data. A.S. performed and analyzed the fluorescence measurements. E.H., A.S., M.S., E.B., B.N. and L.F. discussed the results and implications and commented on the manuscript at all stages.